\documentclass{article}
\usepackage{spconf,graphicx}
\usepackage{mathtools,amssymb,amsthm,amsmath,stmaryrd}

\usepackage{enumitem}
\setlist{nosep, leftmargin=14pt}

\usepackage{mwe,xcolor} 

\title{Implicit Neural Representations for end-to-end PET reconstruction}
\name{\begin{tabular}{c}
Younès Moussaoui$^{1,}$$^{2}$, Diana Mateus$^{1}$, Nasrin Taheri $^{2,}$$^{3}$, Saïd Moussaoui$^{1}$, 
\\
Thomas Carlier$^{2,}$$^{3}$,
Simon Stute $^{2,}$$^{3}$ 
\end{tabular}
\thanks{}
}
\address{$^{1}$Nantes Université, École Centrale Nantes, LS2N, CNRS, UMR 6004, F-44300 Nantes, France
       \\
       $^{2}$Dépt. de Médecine Nucléaire, Centre Hospitalier Universitaire de Nantes, F-44000 Nantes, France
       \\
       $^{3}$CRCI2NA, INSERM, CNRS, Université d'Angers, Université de Nantes, F-44000 Nantes, France
       }

\begin{document}
%

\maketitle
\begin{abstract}

Implicit neural representations (INRs) have demonstrated strong capabilities in various medical imaging tasks, such as denoising, registration, and segmentation, by representing images as continuous functions, allowing complex details to be captured. For image reconstruction problems, INRs can also reduce artifacts typically introduced by conventional reconstruction algorithms. However, to the best of our knowledge, INRs have not been studied in the context of PET reconstruction. In this paper, we propose an unsupervised PET image reconstruction method based on the implicit SIREN neural network architecture using sinusoidal activation functions. Our method incorporates a forward projection model and a loss function adapted to perform PET image reconstruction directly from sinograms, without the need for large training datasets. The performance of the proposed approach was compared with that of conventional penalized likelihood methods and deep image prior (DIP) based reconstruction using brain phantom data and realistically simulated sinograms. The results show that the INR-based approach can reconstruct high-quality images with a simpler, more efficient model, offering  improvements in PET image reconstruction, particularly in terms of contrast, activity recovery, and relative bias.

\end{abstract}

\begingroup
\renewcommand{\thefootnote}{}
\footnotetext{\textcopyright{} 2025 IEEE. Personal use of this material is permitted. Permission from IEEE must be obtained for all other uses, in any current or future media, including reprinting/republishing this material for advertising or promotional purposes, creating new collective works, for resale or redistribution to servers or lists, or reuse of any copyrighted component of this work in other works.}
\endgroup

\section{Introduction}
\label{intro}

Positron Emission Tomography (PET) is a non-invasive imaging modality widely used in medical diagnostics since it provides detailed metabolic and functional information, and offers complementary insights beyond traditional anatomical imaging. However, the raw data (sinograms) collected by PET scanners requires 
adequate reconstruction algorithms to produce meaningful images. Traditional PET reconstruction methods rely heavily on iterative algorithms such as maximum likelihood expectation maximization (MLEM), which minimizes the discrepancy between the acquired sinogram and the predicted sinogram generated 
according to the acquisition (projection) model applied to the estimated image~\cite{shepp1982maximum}. Although effective, these methods can suffer from slow convergence and noise amplification, particularly in low-count scenarios where the signal-to-noise ratio (SNR) is low. To alleviate these issues, post-processing techniques are often used to enhance the reconstructed image quality~\cite{dutta2013non}. For that purpose, various regularization techniques 
have been proposed and adopted to preserve image details and reduce noise impact, 
through penalized maximum likelihood (PML) methods, such as the block Sequential Regularized Expectation Maximization (BSREM)~\cite{ahn2003globally}.


Recently, several deep learning approaches have emerged for PET image reconstruction. Zhu et al.~\cite{zhu2018image} introduced AUTOMAP, a deep learning-based method that reconstructs high-quality PET images from sinograms, but it requires large datasets for training. This data-driven approach relies on a learned mapping from sinograms to images. Therefore, the adaptation of the resulting model to different acquisition protocols is not trivial.
In contrast to data priors, methods like the deep image prior (DIP) rely on the deep network structure itself as a prior. Originally proposed by Ulyanov et al.~\cite{ulyanov2018deep}, DIP fits the network weights to a single image, thus avoiding the need for training data. DIP has been successfully applied to low-dose PET denoising by leveraging the structure of a U-Net-like network as implicit prior knowledge~\cite{cui2019pet}. DIP relies on the network to parameterize the image representation. In practice, DIP denoising takes advantage of CNNs’ spectral bias, where these networks tend to capture low-frequency image components (such as smooth structures) before high-frequency details (like fine edges or noise), acting as inherent denoisers~\cite{rahaman2019spectral}.DIP can be used either as a post-processing step or integrated into iterative optimization updates for end-to-end reconstruction. Hashimoto et al.~\cite{hashimoto2022pet} extended this scheme by incorporating the forward projection into a unified DIP-based PET reconstruction. End-to-end approaches have the advantage of jointly optimizing both image quality and noise reduction.

Lately, Implicit Neural Representations (INRs), also known as neural fields or coordinate-based neural representations~\cite{mildenhall2021nerf}, have gained attention in image processing mainly induced by their ability to 
represent complex and high-frequency functions. These networks learn continuous representations of functions that can be evaluated at arbitrary coordinates, making them well suited to high-detail tasks such as 3D shape representation, in the context of biomedical images, for registration, segmentation~\cite{khan2022implicit} and reconstruction~\cite{mildenhall2021nerf}. 

In addition to their continuous nature, 
INRs are inherently differentiable, 
allowing them to be seamlessly integrated into end-to-end reconstruction algorithms to optimize both image quality and accuracy. For instance, Sitzmann et al.~\cite{sitzmann2020implicit} introduced sinusoidal representation networks (SIREN), using a periodic activation function to represent high-frequency 
contents, and its efficiency has been proven in the application to medical imaging by CT reconstruction~\cite{lejeune2022titan}.



In this work, we explore the application of implicit neural networks in the context of PET image reconstruction by proposing suitable adaptations. Moreover, based on realistic data sets, the effectiveness of the resulting reconstruction approach is proven by comparing it to conventional methods and to a DIP-based strategy, 
in terms of estimated image content and quality.




\begin{figure*}[!tb]
  \centering
  \includegraphics[width=\textwidth]{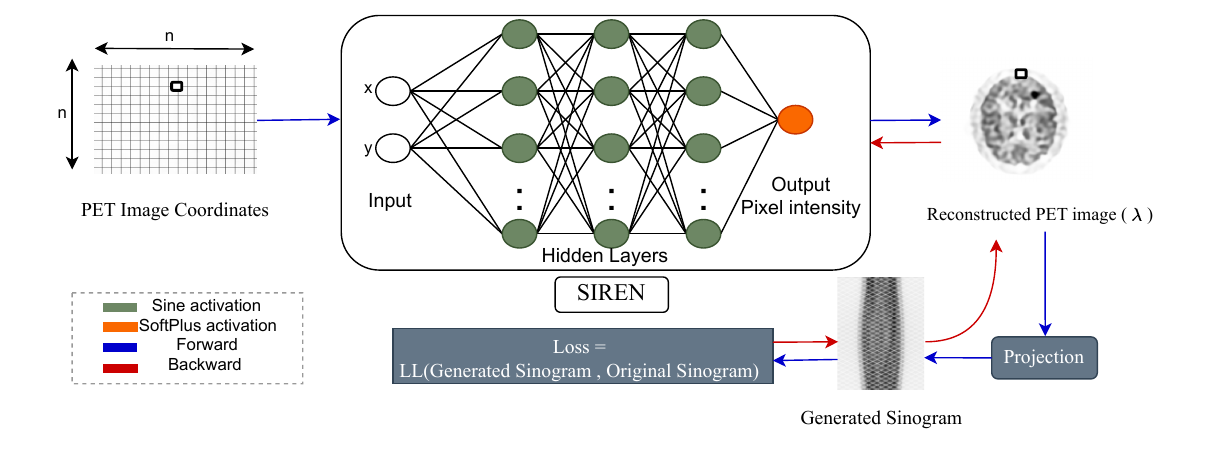}
  \caption{Overview of the proposed pipeline for PET reconstruction using \textit{SIREN}. The model maps the image pixel by pixel to construct the full image. The loss is then computed in the measurement domain between the estimated sinogram obtained by projecting the estimated image and the measured sinogram.}
  \label{model_fig}
\end{figure*}

\section{ Methods}
\label{sec:pagestyle}

\label{methods}


\subsection{End-to-end PET reconstruction}
\label{end-to-end}

The proposed 2D PET image reconstruction framework is illustrated in Fig.~\ref{model_fig}. The reconstructed image, $\lambda \in \mathbb{R}^{n \times n}$  with $n \times n$ the number of voxels, is computed using a deep neural network,
i.e. a SIREN or DIP network, which parameterizes $\lambda$ as follows:
\begin{equation}
\lambda = f_N(\theta\mid v).
\label{eq:im-rep}
\end{equation}
In~\eqref{eq:im-rep},
\( f_N \) represents the function associated to the neural network, 
\( \theta \) refers to the model's weights and \(v\) corresponds to the neural network input, that will be defined later.
%
The reconstructed image \( \lambda \) is obtained by solving the following constrained optimization problem:
\begin{equation}
\label{minproblem}
\min_{\lambda} \left\{-\log p(y \mid \mathcal{P}(\lambda))\right\} 
\quad \text{s.t.} \quad \lambda = f_N(\theta \mid v),
\end{equation}
where \( p(y \mid \mathcal{P}(\lambda)) \) represents the likelihood of the measured data \( y \) given the forward projection \( \mathcal{P}(\lambda) \) of the reconstructed image \( \lambda \). Specifically, the negative log-likelihood used as the loss function is:

\begin{equation}
\label{poisson_loss}
\mathcal{L}_{\text{Poisson}}(y \mid \mathcal{P}(\lambda)) = \sum_i \left( \mathcal{P}(\lambda)_i - y_i \log(\mathcal{P}(\lambda)_i) \right),
\end{equation}

which better suits the Poisson nature of the sinogram data than mean square error (MSE). Here, the forward projection operator is defined as \( \mathcal{P}(\lambda) = \mathbf{A}\lambda + r \), where \( \mathbf{A} \) is the system matrix affected by attenuation and sensitivity, and \( r \) represents background events, including scatter and random coincidences~\cite{stute2015analytical}.

Each iteration $k$ of the optimization involves one forward projection and one backpropagation step: in the forward pass, the image $\lambda_k$ is generated through \( f_N(\theta_k) \), then projected using \( \mathcal{P} \) to obtain a sinogram $y_k$. The loss is computed based on the likelihood between the generated and measured sinogram $y$ as in \eqref{poisson_loss}, and the updated neural network weights $\theta_{k+1}$ are obtained
accordingly using an optimization algorithm (Stochastic gradient descent, Adam, L-BFGS, etc.). This iterative process is repeated until convergence.
%
%
%
\subsection{SIREN for PET reconstruction}
\label{siren_for_pet}
INRs 
model the image intensities 
using a continuous differentiable
function, in contrast to traditional representations that map images to a discrete pixel grid. More precisely, an INR, 
in 2D PET reconstruction,
learns a mapping from a 2D coordinate \( v:= (x, y) \in \mathbb{R}^2 \) to the corresponding pixel intensity value in \( \mathbb{R} \):
\begin{equation}
f_N : \mathbb{R}^2 \to \mathbb{R}, \quad (x, y) \to \lambda(x,y)=f_{N}(\theta\mid (x,y))
\end{equation}
%
Recently, a neural network architecture for INRs that uses sinusoidal activation functions 
was
proposed in \cite{sitzmann2020implicit} and referred to as SIREN. 
Each layer of SIREN
applies an affine transformation followed by a sine function, allowing the network to model continuous and smooth signals. Given 2D coordinates \( (x, y) \), SIREN learns to map these inputs to their corresponding intensities in the image. 

Our proposal to apply SIREN as $f_N$ in~\eqref{minproblem} for PET image reconstruction is motivated by its ability to naturally capture both low and high frequency information, which is critical for accurately reconstructing fine details in PET images. To better adapt the SIREN model to the PET reconstruction framework, we incorporate a SoftPlus activation function in the final layer, which ensures smooth, differentiable, and strictly positive outputs, as negative values are not feasible in this context.
%
%

\subsection{Deep image prior}
\label{dip_for_pet}


An alternative neural network model that can be employed within 
the end-to-end approach is based on a DIP architecture. In this case, \( f_N \) in~\eqref{minproblem} corresponds to a 3D U-Net-inspired design as proposed by Hashimoto et al.~\cite{hashimoto2022pet}, with \( v \) in \eqref{minproblem} initialized as a random image input. This random image serves to guide the reconstruction process gradually, acting as the starting point for the network’s training. Adaptations are also introduced to handle 2D sinograms, with modifications like replacing trilinear upsampling layers with bilinear interpolation, as suggested by Gong et al.~\cite{gong2018pet}, to avoid checkerboard artifacts. Additionally, no skip connections are used to avoid transferring noise from the initial random image too quickly, ensuring a more gradual refinement~\cite{merasli2023influence}. Finally, a SoftPlus activation function is applied after the last layer to ensure non-negative outputs, providing a smooth, differentiable solution that avoids the "dead neurons" problem~\cite{nag2023serf}.


\section{Experiments}

\subsection{Data and PET projection}

To assess the performance of the PET reconstruction, a 2D brain phantom was generated using the BrainWeb database \cite{cocosco1997brainweb}. The reference image from BrainWeb was used to construct a piecewise uniform PET ground truth (GT) that replicates a typical Fluorodeoxyglucose (FDG) brain distribution with standardized uptake values (SUVs) assigned as follows: 6 for gray matter, 2 for white matter, 0.5 for cerebrospinal fluid (CSF), 0.2 for bone, and 1 for other tissues. Additionally, a uniform circular tumor, with SUV = 10 was manually inserted into the white matter in the PET images. The phantom was sampled into 160 x 160 pixels of 2 mm x 2 mm.
Synthetic sinograms are obtained by 2D PET analytical simulations conducted using the software described in \cite{stute2015analytical}. These simulations incorporate the geometry of the Siemens Biograph mMR scanner and include attenuation, efficiency, 35\% random events, 30\% scattered events, and a total of 3.5 million prompts.

\subsection{Image Quality Metrics}

The peak signal-to-noise ratio (PSNR) and the mean structural similarity index measure (MSSIM) are used to assess the reconstructed image quality within the brain. Additionally, the activity recovery (AR) of a hot tumor region is defined as $\text{AR}_h=\hat {\bar \lambda}_h/{\bar \lambda}_h$, where $\hat {\bar \lambda}_h$ and ${\bar \lambda}_h$ represent the mean activity of the hot region in the reconstructed and ground-truth images, respectively. The relative bias (RB) assesses cold region recovery, defined by ventricles filled with cerebrospinal fluid as $
    \text{RB}_c = \big({\hat {\bar \lambda}_c} - {\bar \lambda}_c\big)/ {\bar \lambda}_c$
where $\hat {\bar \lambda}_c$ and ${\bar \lambda}_c$ represent the mean activity of the cold region in the reconstructed and ground-truth images, respectively. Finally, image roughness (IR), computed in the white matter of the brain, measures the standard deviation of voxel values in the phantom background from their mean value:
\begin{equation}
    \text{IR} = \frac{\sqrt{\frac{1}{L-1} \sum_{i \in ROI} (\hat{\lambda}_{r,i} - \hat{\bar{\lambda}}_{r,ROI})^2}}{\hat{\bar{\lambda}}_{r}}
\end{equation}

where $\hat{\lambda}_{r,i}$ is the activity of white matter, $\hat{\bar{\lambda}}_r$ is its mean value, and $L$ is the number of voxels in the region of interest.

\subsection{Experimental Settings}

For all experiments, the optimization is performed using the a quasi-Newton method (the L-BFGS optimizer), with a learning rate of 1 for SIREN and 1.5 for DIP. This approach showed more stable convergence and superior performance in terms of computational time compared to first-order gradient descent algorithms. The SIREN model consists of 5 neural layers, with 256 features per layer, resulting in a total of 329 985 parameters. The model uses sine activations parameterized by  \( \omega_0 = 25 \), The weights are initialized to ensure that the variance of the activations remains consistent across layers, helping to stabilize the training process. The DIP model contains 382 473 parameters, initialized from random Gaussian distribution, with zero mean and unit standard deviation. The BSREM algorithm is applied using the CASToR framework~\cite{merlin2018castor} with a quadratic penalty, and a manually chosen regularization parameter $\beta=0.355$ to achieve optimal PSNR and SSIM between the reconstructed and ground truth image. 


\section{RESULTS \& DISCUSSION}
\label{result_discussion}

\begin{figure}[t]
  \centering
  \includegraphics[width=\columnwidth]{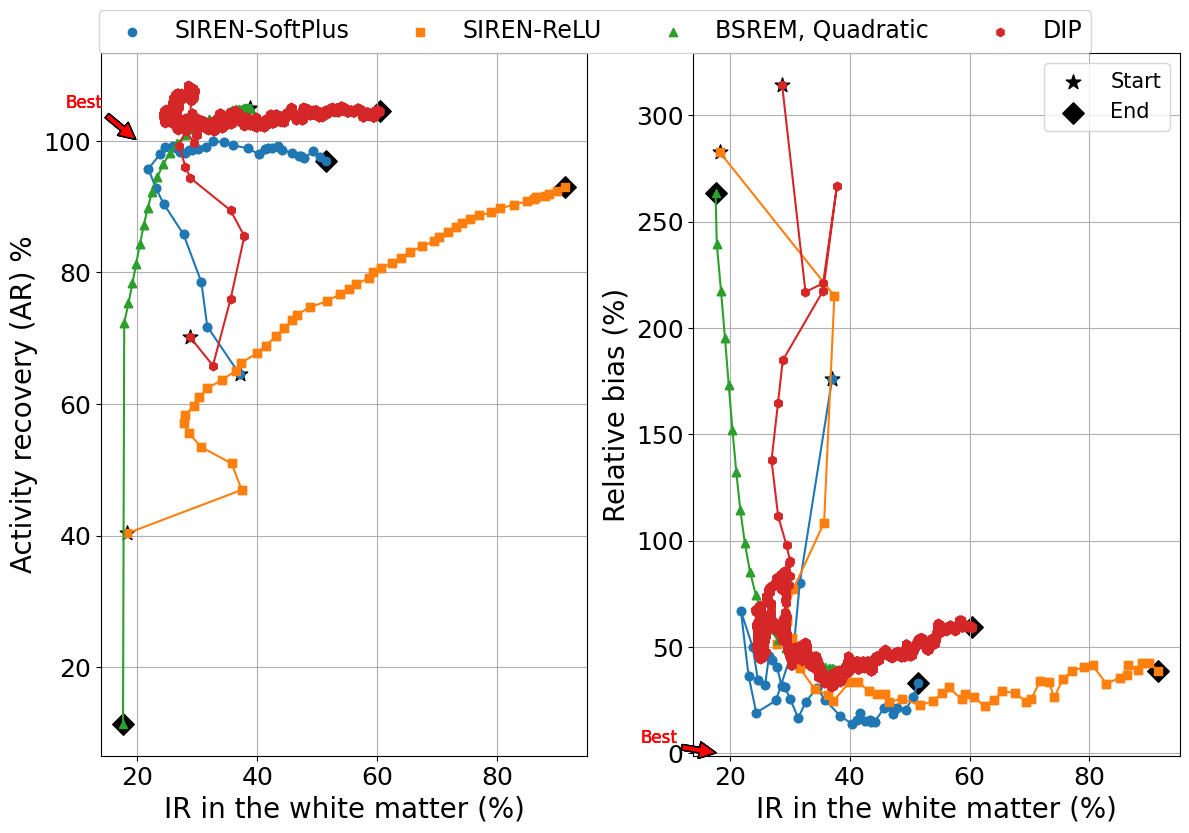}
 \caption{Image Roughness (IR) versus Activity Recovery and Relative Bias, plotted for each iteration for SIREN and DIP, and for different regularization strength in the case of BSREM with quadratic penalty.}
  \label{fig_siren_dip_brsem}
\end{figure}

Fig. \ref{fig_siren_dip_brsem} shows the performance of SIREN in terms of $\text{AR}_h$ and $\text{RB}_c$ using ReLU and SoftPlus activation functions, compared to DIP and BSREM with quadratic penalty. The plot shows results for each iteration in the case of DIP and SIREN, and for a set of regularization parameters in the case of BSREM. We see that SIREN with SoftPlus was able to reach an optimal AR$_h$ value of 100\% in very few iterations, with significantly less roughness than DIP and BSREM. Similarly, SIREN achieved the lowest relative bias of 23\%, with low image roughness compared to both DIP and BSREM.

\begin{figure}[t]
  \centering
  \includegraphics[width=\columnwidth]{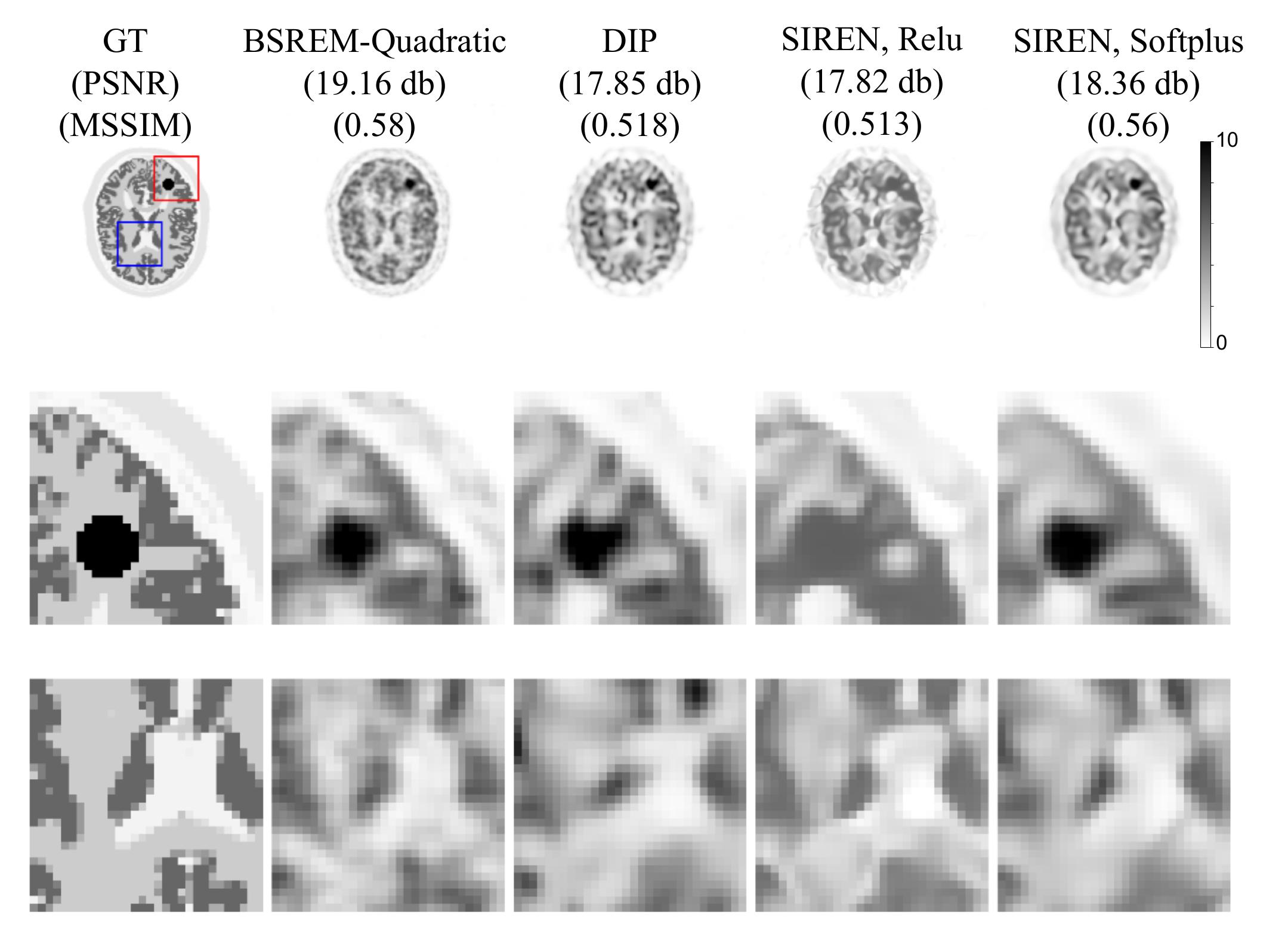}
 \caption{Reconstructed images using BSREM with quadratic penalty, DIP and SIREN. }
  \label{figure_recon}
\end{figure}

Fig. \ref{figure_recon} shows reconstructed images using the compared algorithms. The BSREM reconstruction is shown with the best regularization weight, giving the highest PSNR. The SIREN and DIP reconstructions are shown at the iterations that give the best PSNR. The figure shows that BSREM, with its optimal values, preserves more detail than the other methods, although some blurring is still present. DIP suffers from increased blurring and loss of finer details. SIREN with ReLU, while providing slightly better contrast, struggles to recover the hot region. In contrast, SIREN with SoftPlus strikes a balance between DIP and BSREM, preserving more detail than DIP and offering better contrast than the other methods, as seen in the zoomed-in region of the phantom. 
It is important to note that SIREN with SoftPlus performs better in terms of PET-specific metrics such as $\text{AR}_h$, $\text{RB}_c$, and IR.

In terms of computing time, SIREN achieved the best reconstruction in 20 minutes, compared to 67 minutes for DIP. Each iteration for SIREN took approximately 2 minutes, and only 10 iterations were required to reach the best image quality. In contrast, DIP took approximately 8.5 seconds per iteration but required 475 iterations to yield optimal results. For comparison, one BSREM reconstruction required 7.6 seconds per iteration, but the regularization parameters must be carefully chosen—a process that is not straightforward in practice. These results highlight SIREN's ability to achieve high-quality reconstructions faster and with fewer iterations than DIP.

The proposed PET reconstruction method builds on the approach of Hashimoto et al., but improves upon it by replacing the DIP architecture with a SIREN model. This substitution aims to further reduce biases, including artifacts introduced by the data, and focuses on optimizing reconstruction directly from the measured projection data, scatter and random corrections, and attenuation correction. Normalization is also integrated directly into the reconstruction process to improve consistency.

\section{Conclusion}
\label{sec:conclusion}
In this study, we introduced an unsupervised method for PET image reconstruction by taking advantage of the SIREN implicit neural network architecture.
Inspired by Hashimoto et al, we used a complete pipeline for end-to-end reconstruction, including a forward projection model and a loss function suitable for PET reconstruction directly from sinograms. Comparative evaluations using brain phantom data and realistically simulated sinograms show that the INR-based method outperforms traditional penalized likelihood techniques and DIP-based reconstruction in PET-specific metrics. The results show that our approach improves image contrast and effectively captures fine details. This study highlights the potential of INR to be applied to end-to-end PET reconstruction and to improve overall reconstruction performance. Future work includes extension to 3D PET reconstruction on real data and the automatic fine tuning of the algorithms.


\section{Acknowledgments}
\label{sec:acknowledgments}
This work has been partly funded by the French "Programme d'Investissement d'Avenir" (ANR-16-IDEX-0007) and region "Pays de la Loire" through their support to I-Site NExT, as well as by Siemens Healthineers, the industrial partner of the NExT research industrial chair IMRAM.

\bibliographystyle{IEEEbib}
\bibliography{strings,refs}

\end{document}